# Schwarzschild Field of a Proper Time Oscillator

**Hou Yau**

FDNL Research, 91 Park Manor Drive, Daly City, CA 94015, USA; hyau@fdnresearch.us

**Abstract:** In this paper, we show that an oscillator in proper time can mimic a point mass at rest in general relativity. The spacetime outside this proper time oscillator is static and satisfies the Schwarzschild solution.

**Keywords:** Schwarzschild field; space and time symmetry; proper time oscillator

## 1. Introduction

Nature has a preference for symmetry [1,2]. In classical mechanics, oscillations are considered only in the spatial dimensions. On the other hand, the structure of spacetime is dynamical. Time and space are to be treated on an equal footing in general relativity. Therefore, in theory, we can construct an oscillator that has oscillation in time, if the symmetry between time and space prevails. This oscillator is an analogy of the classical system, except that the oscillation is in time and not in space.

The results attained in ref. [3] lead us to the study of an oscillator in time. As shown in general relativity, the Schwarzschild metric has an extremely simple form which can be expressed as

$$ds^2 = [1 - v(r)^2]dt^2 - [1 - v(r)^2]^{-1}dr^2 - r^2 d\Omega^2, \tag{1}$$

where $v(r) = -(2GM/r)^{1/2}$. It so happens that $v(r)$ is also the free-falling velocity in the gravitational field. The diagonal time component ($g_{tt}$) of the metric is inversely proportional to its diagonal radial component ($g_{rr}$) [4]. Because of this simple form, many attempts have been made to reconcile the metric, by invoking the reciprocity of time dilation and length contraction for a moving object with velocity $v(r)$ [5–10]. These results have led to the erroneous belief that it is possible to derive the Schwarzschild metric by insinuating the concepts of the equivalence principle, special relativity, and Newtonian gravity into the formulation, without referencing Einstein's field equations.

The analyses carried out by Schild [11], Rindler [12], Sacks [13], Gruber [14], and Kassner [15,16] have affirmed the result that there is no simple derivation of the Schwarzschild metric. To illustrate this point, let us follow the arguments made in [14]. In their analysis, Gruber et al. first studied the total mass-energy of a free-falling particle as measured by a stationary observer at a coordinate $r$. With this piece of information, the diagonal component $g_{tt}(r)$ can be inferred from a gedanken experiment with photons. This solution is straightforward and can be derived directly from special relativity. On the other hand, the determination of the *a priori* diagonal component $g_{rr}(r)$ requires the knowledge of a second gravitational property. In this case, Gruber et al. have chosen to specify a "gravitational force law", $f(r) = -d^2r/d\tau^2$, which is a relativistic palatable analogy to the Newtonian radial acceleration with $\tau$ being the proper time measured by the free-falling observer. In Newtonian theory, $v(r) = -(2GM/r)^{1/2}$ and $f(r) = GM/r^2$ for a gravitating body with mass $M$. With both $v(r)$ and $f(r)$ known, the diagonal metric components computed are $g_{tt} = 1 - 2GM/rc^2$ and $g_{rr} = [1 - 2GM/rc^2]^{-1}$. Interestingly, these results are already the exact Schwarzschild solutions. However, when we extend our considerations to relativistic gravity, we expect that $v(r)$ and $f(r)$ will have to be corrected to capture the relativistic gravitational effects. In order to determine these corrections, we will have to make use of information from other sources such as measurements from experiments or theory other than the Newtonian gravity. As a consequence, it is insufficient to determine the Schwarzschild solution only from the equivalence principle, special relativity, and Newtonian gravity without referencing Einstein's field equations.

It is a mere coincidence that the application of time dilation and length contraction can come up with a metric that looks like the correct one. Einstein's field equations cannot be neglected in the formulation. Although it has been vindicated repeatedly that the application of time dilation and length contraction is not sufficient to derive the Schwarzschild metric, the results that can be engendered from this approach are rather appealing. Instead of attempting to derive the Schwarzschild metric, because it is problematic to do so without referencing the field equations, there is another constructive way to adopt the prior application in the gravitational theory.

In [3], we demonstrated that a spherical thin shell with fictitious radial velocity can replicate the gravitational effects of a massive spherical thin shell. The fictitious velocity is analogous to the free-falling velocity except that it is applied only on the surface of a thin shell. Its hypothetical effects on time and distance measurements are adopted to define the spacetime metric on the surface of a time-like hypersurface. Analogously to introducing a fictitious force to describe gravity, we can utilize the fictitious radial velocity to replicate the gravitational effects of a thin shell. The external spacetime outside the thin shell is static and satisfies the Schwarzschild solution for a spherically symmetric mass. According to Birkhoff's theorem [17,18], the thin shell can be contracted to an infinitesimal radius. Based on the time translational symmetry of the system, the same results can be applied to a thin shell with fictitious radial oscillations. As long as the equivalent mass of the shell remains constant during the contraction, the spacetime geometry outside will not be affected. This thin shell with infinitesimal radius is the size of a point mass. In this paper, we show that these fictitious radial oscillations are the products of a proper time oscillator.

The paper is organized in the following manner: In Section 2, we define the basic properties of a proper time oscillator. In Sections 3 and 4, we develop a Lorentz covariant plane wave with vibrations in time and space, which is applied to the Fourier decomposition of the proper time oscillation. In Section 5, we show that the proper time oscillator is encased by fictitious radial oscillations. In Sections 6 and 7, the properties of the fictitious radial oscillations are further elaborated. In Section 8, we investigate the spacetime geometry outside a thin shell with a finite radius that has fictitious oscillations in the radial direction. Based on Birkoff's theorem, the thin shell can be contracted to an infinitesimal radius which is the same thin shell as the one obtained in Section 4. As summarized in the last section, the proper time oscillator has properties that can mimic a stationary point mass in general relativity.

## 2. Proper Time Oscillator

Consider a coordinate system $(t, \mathbf{x})$. The coordinate time $t$ is measured by the clock of a stationary observer $O$ located at spatial infinity. In a Minkowski spacetime, we can synchronize all the stationary clocks with the clock of $O$. Instead of considering a perfectly flat spacetime, let us assume time $\mathring{t}_f$, as observed at the origin of the spatial coordinates $\mathbf{x}_0$, is oscillating with angular frequency $\omega_0$ and amplitude $\mathring{T}_0 = 1/\omega_0$, i.e.,

$$\mathring{t}_f(t, \mathbf{x}_0) = t - \mathring{T}_0 \sin(\omega_0 t). \tag{2}$$

In addition, this proper time oscillator is stationary at $\mathbf{x}_0$,

$$\mathring{\mathbf{x}}_f(t, \mathbf{x}_0) = \mathbf{x}_0, \tag{3}$$

where $\mathring{\mathbf{x}}_f$ is the observed spatial location of the oscillator. Therefore, time $\mathring{t}_f$ observed at the origin is different from time $t$ observed at spatial infinity. Apart from this proper time oscillator, there is no oscillation in time elsewhere.

From Equation (2), the relative rate of time is,

$$\frac{\partial \mathring{t}_f(t, \mathbf{x}_0)}{\partial t} = 1 - \cos(\omega_0 t). \tag{4}$$

Not only is the average of this rate of time 1, its value is bounded between 0 and 2, which is positive. Therefore, time measured at the origin can move only forward. It cannot go back to its past. For a high-frequency oscillator, time observed at the origin will appear to travel along a time-like geodesic if the measuring clock is not sensitive enough to detect the oscillation. We note that the sinuating rate of time is not the result of relative motion. The proper time oscillator is stationary relative to observer $O$ at spatial infinity.

The proper time oscillation is a part of the spacetime geometry. Outside the oscillator, the region is a vacuum spacetime. There is no temporal vibration outside the proper time oscillator, i.e., $\mathring{t}_f(t, \mathbf{x}) = t$ when $\mathbf{x} \neq \mathbf{x}_0$. The temporal vibrations at and around the oscillator can be expressed as

$$\mathring{t}_f(t, \mathbf{x}) = t - \frac{\Pi(\mathbf{x}) \sin(\omega_0 t)}{\omega_0} = t + \mathring{\xi}_t(t, \mathbf{x}), \tag{5}$$

where

$$\mathring{\xi}_t(t, \mathbf{x}) = -\frac{\Pi(\mathbf{x})}{\omega_0} \sin(\omega_0 t), \tag{6}$$

and

$$\Pi(\mathbf{x}) = 0 \text{ if } |\mathbf{x}| \geq \epsilon/2, \tag{7}$$

$$\Pi(\mathbf{x}) = 1 \text{ if } |\mathbf{x}| < \epsilon/2. \tag{8}$$

$\Pi(\mathbf{x})$ is a pulse with width $\epsilon \to 0$.

## 3. Lorentz Covariant Plane Wave

As a wave with vibration in time, $\mathring{\xi}_t$ from Equation (6) can be decomposed into Fourier series of plane waves $\xi_{t\mathbf{k}} = -iT_{\mathbf{k}} e^{i(\mathbf{k}\cdot\mathbf{x} - \omega t)}$, where $T_{\mathbf{k}}$ is an amplitude for the temporal vibrations. However, in a relativistic theory, $\xi_{t\mathbf{k}}$ can only be the 0-component of a Lorentz covariant plane wave. It is necessary to include a spatial component for the Lorentz covariant plane wave that has vibrations in space, i.e., $\xi_{\mathbf{x}\mathbf{k}} = -i\mathbf{X}_{\mathbf{k}} e^{i(\mathbf{k}\cdot\mathbf{x} - \omega t)}$, where $\mathbf{X}_{\mathbf{k}}$ is an amplitude for the spatial vibrations. A Lorentz covariant plane wave shall have vibrations in both the temporal and spatial directions.

To study the vibrations that take place inside this Lorentz covariant plane wave, we adopt a convention similar to the Lagrangian formulation in wave mechanics. Here, $t_d(t, \mathbf{x})$ and $\mathbf{x}_d(t, \mathbf{x})$ are defined as the differences of the time measurement and spatial location from the undisturbed state labeled $(t, \mathbf{x})$. In the Lagrangian formulation, $\mathbf{x}_d(t, \mathbf{x})$ does not tell us the spatial displacement of an observer, which, at time $t$, has coordinate $\mathbf{x}$, but rather, as the displacement of an observer, which has coordinate $\mathbf{x}$ in the undisturbed condition. Similarly, $t_d(t, \mathbf{x})$ is the difference in time, measured by an observer originally from $\mathbf{x}$, relative to the coordinate time $t$ . An observer originally at $\mathbf{x}$ will be displaced to $\mathbf{x}_f = \mathbf{x} + \mathbf{x}_d$, and measures a time $t_f = t + t_d$, instead of the time $t$ at spatial infinity. The coordinate time $t$ is used as a reference for measuring the temporal vibrations inside the wave.

Let us first consider a plane wave with only temporal vibrations as observed in an inertial frame $O'$. We will define the wave's temporal amplitude $T_0$ as the maximum difference between time $t'_f$ observed inside the wave and time $t'$, observed outside the wave by an inertial observer. We may then write

$$t'_f = t' - T_0 \sin(\omega_0 t') = t' + t'_d = t' + \mathrm{Re}(\xi'_t), \tag{9}$$

$$\mathbf{x}'_f = \mathbf{x}', \tag{10}$$

where

$$t'_d = \mathrm{Re}(\xi'_t) = -T_0 \sin(\omega_0 t'), \tag{11}$$

$$\xi'_t = -iT_0 e^{-i\omega_0 t'}. \tag{12}$$

This plane wave has no vibration in space.

The background coordinates $(t', \mathbf{x}')$ of inertial frame $O'$ can be Lorentz-transformed from the background coordinates $(t, \mathbf{x})$ for the flat spacetime in another frame of reference $O$,

$$t' = \gamma(t - \mathbf{v} \cdot \mathbf{x}), \tag{13}$$

$$\mathbf{x}' = \gamma(\mathbf{x} - \mathbf{v}t), \tag{14}$$

where

$$\gamma = (1 - |\mathbf{v}|^2)^{-1/2}. \tag{15}$$

We assumed that $O'$ is travelling with a velocity $\mathbf{v}$ relative to $O$. Similarly, the displaced coordinates $(t'_f, \mathbf{x}'_f)$ can be Lorentz-transformed to the displaced coordinates $(t_f, \mathbf{x}_f)$ as observed in frame $O$,

$$t_f = \gamma(t'_f + \mathbf{v} \cdot \mathbf{x}'_f), \tag{16}$$

$$\mathbf{x}_f = \gamma(\mathbf{x}'_f + \mathbf{v}t'_f). \tag{17}$$

Substitute Equations (9), (10), (13), and (14) into Equations (16) and (17), the displaced coordinates $(t_f, \mathbf{x}_f)$ are,

$$t_f = t + T\sin(\mathbf{k} \cdot \mathbf{x} - \omega t) = t + t_d = t + \mathrm{Re}(\xi_t), \tag{18}$$

$$\mathbf{x}_f = \mathbf{x} + \mathbf{X}\sin(\mathbf{k} \cdot \mathbf{x} - \omega t) = \mathbf{x} + \mathbf{x}_d = \mathbf{x} + \mathrm{Re}(\xi_{\mathbf{x}}), \tag{19}$$

where

$$t_d = \mathrm{Re}(\xi_t) = T\sin(\mathbf{k} \cdot \mathbf{x} - \omega t), \tag{20}$$

$$\mathbf{x}_d = \mathrm{Re}(\xi_{\mathbf{x}}) = \mathbf{X}\sin(\mathbf{k} \cdot \mathbf{x} - \omega t), \tag{21}$$

$$\xi_t = -iTe^{i(\mathbf{k} \cdot \mathbf{x} - \omega t)}, \tag{22}$$

$$\xi_{\mathbf{x}} = -i\mathbf{X}e^{i(\mathbf{k} \cdot \mathbf{x} - \omega t)}, \tag{23}$$

$$T = (\omega/\omega_0)T_0, \tag{24}$$

$$\mathbf{X} = (\mathbf{k}/\omega_0)T_0. \tag{25}$$

Amplitude $\mathbf{X}$ is the maximum displacement of the wave from its undisturbed coordinate $\mathbf{x}$, and amplitude $T$ is its maximum displacement from time $t$. The proper time displacement $T_0$ can be seen as a Lorentz transformation of a 4-displacement vector: $(T_0, 0, 0, 0) \to (T, \mathbf{X})$ where $T^2 = T_0^2 + |\mathbf{X}|^2$. The amplitude of the plane wave is a 4-vector.

Inside a plane wave, the temporal and spatial coordinates are displaced from the undisturbed state. We can further summarize these vibrations with a single function,

$$\xi = \frac{T_0}{\omega_0} e^{i(\mathbf{k} \cdot \mathbf{x} - \omega t)}. \tag{26}$$

The vibrations $\xi_t$ and $\xi_{\mathbf{x}}$ from Equations (22) and (23) can be written as:

$$\xi_t = \partial_0 \xi, \tag{27}$$

$$\xi_{\mathbf{x}} = -\nabla \xi. \tag{28}$$

In the rest of this paper, we consider $T_0$, $T$ and $\mathbf{X}$ as complex amplitudes.

## 4. Fourier Decomposition of the Proper Time Oscillation

Let us define a function,

$$\mathring{\xi}(t,\mathbf{x}) = \frac{\Pi(\mathbf{x})}{\omega_0^2}\cos(\omega_0 t), \tag{29}$$

which can be decomposed into a Fourier series of plane waves $\xi_{\mathbf{k}}$ from Equation (26), and their complex conjugates $\xi_{\mathbf{k}}^*$. Since the superposition is linear, we can also apply Equations (27) and (28) to obtain the corresponding vibrations in time and space $(\mathring{\xi}_t, \mathring{\xi}_{\mathbf{x}})$ of the superposed function $\mathring{\xi}$, i.e.,

$$\mathring{\xi}_t = \partial_0 \mathring{\xi}, \tag{30}$$

$$\mathring{\xi}_{\mathbf{x}} = -\nabla \mathring{\xi}. \tag{31}$$

Hence, $\mathring{\xi}_t$ from Equation (6) can be obtained from Equations (29) and (30),

$$\mathring{\xi}_t = \partial_0 \mathring{\xi} = -\frac{\Pi(\mathbf{x})}{\omega_0}\sin(\omega_0 t). \tag{32}$$

On the other hand, Equation (31) implies that there are spatial oscillations in addition to the temporal oscillation at $\mathbf{x}_0$. The origins of these additional oscillations stem from the spatial components of the Lorentz covariant plane waves. Since the system that we are considering is spherically symmetric, we can switch to a spherical coordinate system. The oscillations in space are described by $\mathring{\xi}_r$,

$$\mathring{r}_f(t,r) = r + \mathring{\xi}_r(t,r), \tag{33}$$

where

$$\mathring{\xi}_r(t,r) = -\frac{\partial \mathring{\xi}}{\partial r} = -\frac{\Pi'(r)}{\omega_0^2}\cos(\omega_0 t). \tag{34}$$

$\Pi'(r)$ denotes the derivative of $\Pi(r)$ with respect to r, such that

$$\Pi'(r) = 0 \text{ if } r \neq \epsilon/2, \tag{35}$$

$$\Pi'(r) = -\infty \text{ if } r = \epsilon/2. \tag{36}$$

Therefore, apart from the proper time oscillation at $r = 0$, there are oscillations in the radial direction about $r = \epsilon/2$. The radial oscillations about $r = \epsilon/2$ are revealed only after we study the Fourier decomposition of the proper time oscillation.

## 5. Fictitious Radial Oscillation with Infinite Amplitude

We can summarize our results from Section 4 as follow:

At $r = 0$,

$$\mathring{t}_f(t,0) = t - \frac{\sin(\omega_0 t)}{\omega_0}, \tag{37}$$

$$\mathring{r}_f(t,0) = 0. \tag{38}$$

At $r = \epsilon/2 \to 0$,

$$\mathring{t}_f(t,\epsilon/2) = t, \tag{39}$$

$$\mathring{r}_f(t,\epsilon/2) = \epsilon/2 + \Re_\infty \cos(\omega_0 t) \text{ with } \Re_\infty \to \infty. \tag{40}$$

The system is spherically symmetric with oscillations in the temporal and radial directions only. From Equation (38), the proper time oscillator is stationary at the origin of the spatial coordinate, $r = 0$. It has no vibration in space where the oscillator is located. On the other hand, the region outside the proper time oscillator is a vacuum spacetime that is source-free. There are no vibrations in this vacuum spacetime except about a thin shell with radius $r = \epsilon/2 \to 0$, as shown in Equation (40).

Let us look at this system in more detail. As shown, the system has two oscillating components: the proper time oscillator at $r = 0$ and the radial oscillations about $r = \epsilon/2$. They are simple harmonic oscillators. Based on our knowledge about simple harmonic oscillating systems, their total energy shall be conserved over time. Therefore, we expect that the above system as a whole will have a symmetry under time translation as stipulated by the Noether's theorem (The assumption of time translational symmetry will be falsified if we cannot obtain the Schwarzschild field solution from the proper time and radial oscillations. As we have learned from general relativity, the solution of Einstein's field equations for static spherically symmetric system is a Schwarzschild field.). Moreover, the spacetime outside the proper time oscillator is a vacuum. As shown in Equation (39), a radial oscillation about $r = \epsilon/2$ does not have oscillation in time.

Matter cannot oscillate in space with an infinite amplitude. This would violate the principles of relativity by allowing superluminal transfer of energy. From Equation (40), the instantaneous velocity of the radial oscillation is

$$\mathring{v}_f(t, \epsilon/2) = \frac{\partial}{\partial t}\mathring{r}_f(t, \epsilon/2) = -\Re_\infty \omega_0 \sin(\omega_0 t), \tag{41}$$

which can exceed the speed of light. The transportation of an observer by the radial oscillation through space is forbidden by the principles of relativity. The radial oscillation, therefore, cannot be interpreted as a vibration that can carry an observer through space. Instead, we study the effects of this radial oscillation on an observer that is stationary at $r = \epsilon/2$.

Any effects resulting from the radial oscillations are negligible at spatial infinity where the spacetime is considered as flat. An observer $O$, stationary at spatial infinity, is an inertial observer which is used as the reference for our study. In a Minkowski spacetime, the clock of any stationary observer can be synchronized with the clock of $O$. However, this is not the case for an observer $O_+$ stationary at $r = \epsilon/2$. As shown in Equations (39) and (40), it is the clock of a fictitious observer $\underline{O}$ oscillating about $r = \epsilon/2$ that synchronizes with the clock of $O$. As discussed, this oscillation cannot have physical vibration through space since it will allow superluminal transfer of energy. We shall consider its effects at $r = \epsilon/2$, as if $O_+$ is oscillating in a fictitious frame of $\underline{O}$. In this fictitious frame, $\underline{O}$ is an inertial observer. An observer $O_+$, on the shell with radius $r = \epsilon/2$, will have an oscillation $\underline{r}_f$ relative to the fictitious inertial observer $\underline{O}$, i.e.,

$$\underline{r}_f(t, \epsilon/2) = -\Re_\infty \cos(\omega_0 t), \tag{42}$$

$$\underline{v}_f(t, \epsilon/2) = \frac{\partial}{\partial t}\underline{r}_f(t, \epsilon/2) = \Re_\infty \omega_0 \sin(\omega_0 t), \tag{43}$$

where $\underline{v}_f$ is the instantaneous velocity of the fictitious radial oscillation. Therefore, $O_+$ is under the constant effects of a fictitious oscillation while remaining at rest relative to $O$ at spatial infinity. These fictitious radial oscillations are used to study the spacetime geometry outside the proper time oscillator.

Taking the thin shell with fictitious radial oscillations as a part of the spacetime geometry, its properties are different from those of the assumed flat spacetime at spatial infinity where there is no oscillation. The geometry of spacetime at these two distant locations is different. Therefore, if the spacetime manifold outside the thin shell is smooth and continuous, its structures cannot be flat.

In the previous section, the superposition of the Lorentz covariant plane waves is carried out in a flat spacetime. The result obtained from Equation (33) shows that the proper time oscillator is bounded by a thin shell with fictitious radial oscillations. The radius of this thin shell is infinitesimally small and there are no other oscillations outside. All the oscillations of the system are contained in an infinitesimal region. Therefore, this oscillating system can be localized in a region where spacetime is only locally flat. If this system is placed in an initially flat spacetime, we expect the thin shell will curve the surrounding spacetime as discussed above. As we demonstrate in Section 8, the resulting spacetime geometry satisfies the Schwarzschild solution and can mimic the gravitational field of a point mass in general relativity. The assumption of flat spacetime, when we carry out the superposition of plane waves in the previous section, is locally true at the origin of the spatial coordinates.

## 6. Fictitious Radial Oscillation with Finite Amplitude

Instead of working directly with the fictitious radial oscillations about $r = \epsilon/2$, let us first consider an infinitesimally thin spherical shell $\Sigma$ with finite radius $\check{r}$. Relative to this shell, there are fictitious radial oscillations [3], i.e.,

$$\underline{r}_f(t, \check{r}) = -\check{\Re}\cos(\omega_0 t), \tag{44}$$

$$\underline{v}_f(t, \check{r}) = \frac{\partial}{\partial t}\underline{r}_f(t, \check{r}) = \check{\Re}\omega_0 \sin(\omega_0 t), \tag{45}$$

where $\check{\Re}\omega_0 < 1$. The properties of these fictitious radial oscillations with amplitude $\check{\Re}$ are analogous to those about $r = \epsilon/2$, except the magnitude of the instantaneous velocity $|\underline{v}_f|$ is now less than 1, and the amplitude of oscillation $\check{\Re}$ is finite. Apart from the fictitious oscillations in the radial direction, there are no other oscillations. The spacetime outside the shell is a vacuum and the system is spherically symmetric. We will further assume the system has a time translational symmetry, as expected for a simple harmonic oscillating system.

Analogously to the thin shell with infinitesimal radius, it is the clock of a fictitious observer $\underline{O}$ oscillating about $r = \check{r}$ that synchronizes with the clock of an observer $O$ at spatial infinity. In its fictitious frame, $\underline{O}$ is an inertial observer. An observer $\check{O}$, stationary at $r = \check{r}$, has a fictitious displacement $\underline{r}_f$ and instantaneous velocity $\underline{v}_f$ relative to $\underline{O}$. Although $\check{O}$ is stationary relative to $O$ at spatial infinity, it is under the effects as if $\check{O}$ is oscillating in the fictitious frame of $\underline{O}$. As discussed previously, the fictitious oscillation is not a vibration that carries an observer through space. Instead, their effects on time and distance measurements are used to define the metric on the surface of the thin shell $\Sigma$. In other words, the fictitious oscillations are no vibrations of matter. Only their geometrical effects are considered. Therefore, the thin shell with fictitious oscillations is different from an oscillating thin shell star that has physical motions of matter.

The properties of a moving observer with $|\mathbf{v}| < 1$ are well defined in relativity. These properties can be applied in the fictitious frame of $\underline{O}$. However, apart from the instantaneous velocity $\underline{v}_f$, $\check{O}$ also has a displacement $\underline{r}_f$ relative to the fictitious observer $\underline{O}$. As a simple oscillating system, we expect both the fictitious displacement and its instantaneous velocity can have effects on $\check{O}$. Their combined effects shall remain constant such that the total Hamiltonian of the system is invariant over time. Although the effects of the fictitious displacement are not yet defined, we can obtain the metric induced on the thin shell $\Sigma$ when there is only a fictitious velocity with $|\underline{v}_f| < 1$.

At $t = t_m = \pi/(2\omega_0)$, the fictitious displacement and instantaneous velocity from Equations (44) and (45) are:

$$\underline{r}_f(t_m, \check{r}) = \underline{r}_{fm} = 0, \tag{46}$$

and

$$\underline{v}_f(t_m, \check{r}) = \underline{v}_{fm} = \check{\Re}\omega_0 < 1. \tag{47}$$

$\check{O}$ is traveling with a velocity $\underline{v}_{fm}$ in the fictitious frame with no displacement relative to $\underline{O}$. To obtain the metric at $r = \check{r}$, we need to understand how the clocks and measuring rods carried by $O$ and $\check{O}$ are related at the instant $t = t_m$.

## 7. Measurements on the Thin Shell $\Sigma$

Let us consider two events observed in frame $\check{O}$ at the instant $t = t_m$. The infinitesimal coordinate increments $(dt, dr)$ observed in frame $O$, can be expressed in terms of the infinitesimal coordinate increments $(d\check{t}, d\check{r})$, for the same two events observed in frame $\check{O}$,

$$\begin{bmatrix} dt \\ dr \end{bmatrix} = \begin{bmatrix} \Upsilon^{t}{}_{\check{t}} & \Upsilon^{t}{}_{\check{r}} \\ \Upsilon^{r}{}_{\check{t}} & \Upsilon^{r}{}_{\check{r}} \end{bmatrix} \begin{bmatrix} d\check{t} \\ d\check{r} \end{bmatrix}. \tag{48}$$

In the local frame of $O$, the basis vectors in the temporal and radial directions are orthogonal,

$$\vec{e}_t \cdot \vec{e}_r = 0. \quad (49)$$

The same is true in frame $\check{O}$,

$$\vec{e}_{\check{t}} \cdot \vec{e}_{\check{r}} = 0. \quad (50)$$

As discussed, $\check{O}$ is stationary relative to $O$. The temporal basis vectors in frames $O$ and $\check{O}$ are, therefore, parallel to one another,

$$\vec{e}_{\check{t}} \parallel \vec{e}_t. \quad (51)$$

Similarly, the radial basis vectors of the two frames are also parallel,

$$\vec{e}_{\check{r}} \parallel \vec{e}_r. \quad (52)$$

Under conditions (49), (50), (51) and (52), the transformation matrix $\Upsilon$ is diagonal , i.e.,

$$\begin{bmatrix} dt \\ dr \end{bmatrix} = \begin{bmatrix} \Upsilon^t{}_{\check{t}} & 0 \\ 0 & \Upsilon^r{}_{\check{r}} \end{bmatrix} \begin{bmatrix} d\check{t} \\ d\check{r} \end{bmatrix}. \quad (53)$$

To determine the element $\Upsilon^t{}_{\check{t}}$, we will consider the infinitesimal coordinate increments $(d\check{t}, d\check{r} = 0)$, where $d\check{t}$ is a proper time measured by $\check{O}$. Under a Lorentz transformation, the coordinate increments $(d\underline{t}, d\underline{r})$ observed in the fictitious frame $\underline{O}$ are

$$d\underline{t} = \underline{\gamma} d\check{t}, \quad (54)$$

$$d\underline{r} = \underline{\underline{\gamma} \underline{v}_{fm} d\check{t}}, \quad (55)$$

and

$$\underline{\gamma} = [1 - (\underline{v}_{fm})^2]^{-1/2}. \quad (56)$$

Since the clocks of $O$ and $\underline{O}$ are synchronized, we have

$$dt = d\underline{t} = \underline{\gamma} d\check{t}. \quad (57)$$

Although $\check{O}$ appears to be moving in the fictitious frame of $\underline{O}$, there is no relative movement between $\check{O}$ and $O$. The underlined quantity in Equation (55) is a fictitious displacement that is observed only in $\underline{O}$. The time interval measured by $O$ is lengthened by the effects of the fictitious movement with respect to the measurement made by $\check{O}$. Based on Equations (47) and (57),

$$\Upsilon^t{}_{\check{t}} = \underline{\gamma} = [1 - (\underline{v}_{fm})^2]^{-1/2} = (1 - \Re^2 \omega_0^2)^{-1/2}. \quad (58)$$

To determine the element $\Upsilon^r{}_{\check{r}}$, we will consider another infinitesimal coordinate increments $(d\check{t} = 0, d\check{r})$, where $d\check{r}$ is the length of a rod carried by $\check{O}$. Under a Lorentz transformation, the coordinate increments $(d\underline{t}, d\underline{r})$ observed in the fictitious frame $\underline{O}$ are

$$d\underline{t} = \underline{\underline{\gamma} \underline{v}_{fm} d\check{r}}, \quad (59)$$

$$d\underline{r} = \underline{\gamma} d\check{r}. \quad (60)$$

The moving length $d\underline{l}$ of the rod as observed in fictitious frame $\underline{O}$ is

$$d\underline{l} = d\underline{r} - \underline{\underline{v}_{fm} dt} = \underline{\gamma}^{-1} d\check{r}, \quad (61)$$

where the underlined quantity is the distance traveled by the rod during $d\underline{t}$. Since $O$ and $\underline{O}$ are inertial observers that have their clocks synchronized, they shall measure the same length for the same rod, i.e.,

$$dr = d\underline{l} = \underline{\gamma}^{-1} d\check{r}. \tag{62}$$

On the other hand, there is no relative movement between $O$ and $\check{O}$. Therefore, the underlined quantities in Equations (59) and (61) are fictitious temporal and spatial displacements that are observed only in $\underline{O}$. The length of the rod measured by $O$ is shortened by the effects of the fictitious movement with respect to the measurement made by $\check{O}$. Based on Equations (47) and (62),

$$\Upsilon^{r}{}_{\check{r}} = \underline{\gamma}^{-1} = [1 - (\underline{v}_{fm})^2]^{1/2} = (1 - \check{\Re}^2\omega_0^2)^{1/2}. \tag{63}$$

## 8. Schwarzschild Field

At $t = t_m$, the relationships between the clocks and measuring rods of observers $O$ and $\check{O}$ are shown in Equations (53), (58) and (63) (As discussed in the Introduction, a similar approach was taken in refs. [5–9] to obtain the Schwarzschild field by invoking the time dilation and length contraction associated with a free-falling velocity, $v(r) = -(2m/r)^{1/2}$. However, it has also been demonstrated that Einstein's field equations cannot be neglected in deriving this solution [11–15]. The fictitious velocity studied here is analogous to the free-falling velocity, except that we limited its application only on the surface of a massive spherical thin shell [3].). Here, we extend its implications further. As discussed, the simple harmonic oscillating system has a symmetry under time translation. The effects of the fictitious radial oscillations on $\check{O}$ are constant over time. Under these conditions, we can define a constant

$$\check{I} = \omega_0^2(\underline{r}_f)^2 + (\underline{v}_f)^2 = \check{\Re}^2\omega_0^2, \tag{64}$$

such that Equation (53) becomes

$$\begin{bmatrix} dt \\ dr \end{bmatrix} = \begin{bmatrix} (1-\check{I})^{-1/2} & 0 \\ 0 & (1-\check{I})^{1/2} \end{bmatrix} \begin{bmatrix} d\check{t} \\ d\check{r} \end{bmatrix}. \tag{65}$$

The first and second terms on the right-hand side of Equation (64) are equivalent to the 'potential' and 'kinetic' components of a classical harmonic oscillating system. Their total effects are constant under time transnational symmetry.

The system is spherically symmetric with oscillations only in the radial and temporal directions. In addition, there is no rotational motion. The line element at $r = \check{r}$ for such a spherically symmetric system can be written as [19],

$$ds^2 = g_{tt}(\check{r})dt^2 + 2g_{tr}(\check{r})dtdr + g_{rr}(\check{r})dr^2 - \check{r}^2 d\Omega^2. \tag{66}$$

The coordinate time $t$ is measured by a stationary clock located infinitely far from the source as adopted in Section 2. The radial coordinate $r$ can be defined as the circumference, divided by $2\pi$, of a sphere centered around the shell. The angular coordinates $\theta$ and $\phi$ are the usual polar spherical angular coordinates. This coordinate system is the same as the one adopted in the conventional Schwarzschild field.

From Equation (65), the basis vectors in frames $O$ and $\check{O}$ satisfy the following relationships,

$$\vec{e}_{\check{t}} = \vec{e}_t(1-\check{I})^{1/2}, \tag{67}$$

$$\vec{e}_{\check{r}} = \vec{e}_r(1-\check{I})^{-1/2}, \tag{68}$$

where

$$\vec{e}_t \cdot \vec{e}_t = 1, \tag{69}$$

$$\vec{e}_r \cdot \vec{e}_r = -1, \tag{70}$$

$$\vec{e}_t \cdot \vec{e}_r = 0. \tag{71}$$

Therefore, the line elements at $O$ and $\check{O}$ are different. From Equations (67) and (68),

$$g_{tt}(\check{r}) = \vec{e}_{\check{t}} \cdot \vec{e}_{\check{t}} = (1-\check{I})\vec{e}_t \cdot \vec{e}_t = 1-\check{I}, \tag{72}$$

$$g_{rr}(\check{r}) = \vec{e}_{\check{r}} \cdot \vec{e}_{\check{r}} = (1-\check{I})^{-1}\vec{e}_r \cdot \vec{e}_r = -(1-\check{I})^{-1}, \tag{73}$$

$$g_{tr}(\check{r}) = g_{rt}(\check{r}) = \vec{e}_{\check{t}} \cdot \vec{e}_{\check{r}} = \vec{e}_t \cdot \vec{e}_r = 0. \tag{74}$$

At $r = \check{r}$, the line element from Equation (66) is reduced to

$$ds^2 = [1-\check{I}]dt^2 - [1-\check{I}]^{-1}dr^2 - \check{r}^2 d\Omega^2. \tag{75}$$

If we set,

$$\check{I} = \frac{2m}{\check{r}}, \tag{76}$$

or

$$m = \frac{\check{r}\check{\Re}^2\omega_0^2}{2}, \tag{77}$$

Equation (75) becomes the Schwarzschild line element on the surface of a thin shell with total mass $m$. From general relativity, the vacuum space–time $v^+$ outside this time-like hypersurface is the Schwarzschild spacetime,

$$ds^2 = [1 - \frac{\check{r}\check{\Re}^2\omega_0^2}{r}]dt^2 - [1 - \frac{\check{r}\check{\Re}^2\omega_0^2}{r}]^{-1}dr^2 - r^2 d\Omega^2, \tag{78}$$

which is static with time translational and time reflection symmetries.

The time-like hypersurface $\Sigma$ can be contracted by 'carrying' the fictitious oscillations along geodesics orthogonal to the original surface to a new sphere $\Sigma'$. According to the Birkhoff's theorem [17,18], the external gravitational field of the spherically symmetric vacuum region is static and satisfies the Schwarzschild solution. As long as the equivalent mass $m$ from Equation (77) is kept as a constant during the contraction, the metric and curvature of the external field will remain unaffected. Under this condition, the amplitude of the radial oscillation is

$$\check{\Re} = \sqrt{\frac{2}{\check{r}m}}. \tag{79}$$

As the shell is contracted to a radius $\check{r} = 2m$, the metric from Equation (78) will encounter a coordinate singularity (event horizon). Although the fictitious instantaneous velocity can exceed the speed of light ($\underline{v}_{fm} > 1$ when $\check{r} < 2m$), the fictitious oscillations are not physical vibrations of matter. There is no superluminal transfer of energy by the fictitious oscillations. Only the effects of the fictitious oscillations on an observer are physical. Moreover, the amplitude of fictitious oscillation is well defined until the radius is contracted to $\check{r} = \epsilon/2 \to 0$ (The same is true for all the curvature tensors derived from the metric, which is also well defined until the radius of the shell is contracted to a radius $\check{r} = \epsilon/2 \to 0$.). The shell, therefore, can be contracted even beyond a radius $\check{r} = 2m$, as allowed by Birkhoff's theorem, while maintaining the same Schwarzschild geometry.

From Equation (79), the amplitude of fictitious oscillation is infinitely large ($\check{\Re} \to \infty$) on a thin shell with radius $\check{r} = \epsilon/2 \to 0$. This thin shell with an infinitesimal radius is the same one as derived from the proper time oscillator in Section 4. As a result, the spacetime geometry outside the proper time oscillator shall satisfy the Schwarzschild solution. The external spacetime geometry resulting from the proper time oscillator can, therefore, mimic the gravitational field of a point mass in general relativity.

## 9. Conclusions and Discussions

"Spacetime tells matter how to move; matter tells spacetime how to curve."–John Archibald Wheeler. The gravitational effects of matter on the geometry of spacetime are governed by Einstein field equations. However, the theory lacks a mechanical description of how spacetime is curved by matter. In this paper, we show that an oscillator in proper time can mimic the gravitational field of a point mass. The external spacetime geometry of this proper time oscillator is static and satisfies the Schwarzschild solution. These results can paint a simple picture of how matter can be connected to spacetime. The proper time oscillator exerts fictitious radial oscillations on a thin shell with an infinitesimal radius. This alters the spacetime metric on the surface of the thin shell and curves the surrounding external spacetime. In turn, the curved spacetime tells other matter how to react in the presence of the proper time oscillator. The model can establish a more direct correlation between spacetime and matter that can supplement the general theory.

In the quantum theory of fields, particles are treated as sets of coupled quantum oscillators. These particles, as conjectured by de Broglie, shall possess an internal clock. Interestingly, this hypothesis has only been tested recently, and the experimental data obtained are found to be compatible with the theory [20]. If we assume a particle can remain stationary in space and has proper time oscillation, the angular frequency $\omega_0$ adopted in our formulation can be taken as the frequency of the de Broglie's internal clock. This frequency shall be unique for the proper time oscillator/particle. However, we shall bear in mind that the quantum effects of a particle cannot be neglected. A complete quantum gravity theory is required before we can fully understand whether there are any connections between a proper time oscillator and a quantum particle.

The criterion for a true singularity is geodesic incompleteness [21,22]. A singularity is present when the world line of a freely falling test object cannot be extended past that point. In our model, the spacetime geometry is discontinuous at $r = \epsilon/2$ where any causal geodesic (time-like or null) cannot be extended further. This boundary of the proper time oscillator is a singularity of the system. The fictitious radial oscillations on this boundary are the sources of the Schwarzschild geometry. Although the singularity acts as a boundary for incoming geodesics, the spacetime structure inside this boundary is well defined. The proper time oscillator is cloistered behind the singularity of our model.

**Funding:** This research received no external funding.

**Conflicts of Interest:** The author declares no conflict of interest.